# Intestinal Microecology in Pediatric Surgery-Related Gastrointestinal Diseases: Current Insights and Future Perspectives

Yuqing Wu[a], Suolin Li[a], Lin Liu[a], Xiaoyi Zhang[a], Jiaxun Lv[a], Qinqin Li[a], Yingchao Li[a,*]

[a] *Department of Pediatric Surgery, The Second Hospital of Hebei Medical University, Shijiazhuang, China*

**Category of the manuscript:** Review

**Previous communication:** This paper is not based on previous communication to a society or meeting

**Financial support statement:** This research did not receive any specific grant from funding agencies in the public, commercial, or not-for-profit sectors.

[*] Corresponding author at: 215 Heping West Road, Shijiazhuang City, Hebei Province, China
E-mail address: xicani@126.com.

# Intestinal Microecology in Pediatric Surgery-Related Gastrointestinal Diseases: Current Insights and Future Perspectives

**Abstract**

Intestinal microecology is established from birth and is constantly changing until homeostasis is reached. Intestinal microecology is involved in the immune inflammatory response of the intestine and regulates the intestinal barrier function. The imbalance of intestinal microecology is closely related to the occurrence and development of digestive system diseases. In some gastrointestinal diseases related to pediatric surgery, intestinal microecology and its metabolites undergo a series of changes, which can provide a certain basis for the diagnosis of diseases. The continuous development of microecological agents and fecal microbiota transplantation technology has provided a new means for its clinical treatment. We review the relationship between pathogenesis, diagnosis and treatment of pediatric surgery-related gastrointestinal diseases and intestinal microecology, in order to provide new ideas and methods for clinical diagnosis, treatment and research.

**Level of evidence**: This is a review study, hence not applicable

**Abbreviations**: CD, Crohn's disease; FXR, farnesol X receptor; GPCRs, G protein-coupled receptors; HAEC, Hirschsprung-associated enterocolitis; HDAC, histone deacetylase activity; HSCR, Hirschsprung's disease; IBD, inflammatory bowel disease; ICIs, immune checkpoint inhibitors; IMC, immune-mediated colitis; LGG, Lactobacillus rhamnosus GG; NEC, necrotizing enterocolitis; NLRs, nod-like receptors; PXR pregnane X receptor; SBS, short bowel syndrome; SCFAs, short-chain fatty acids; TLRs, Toll-like receptors; UC, ulcerative colitis; VOCs, volatile organic compounds

# 1 Background

Intestinal microecology refers to an ecosystem resulting from the interaction between microorganisms inhabiting the human gut and body. This ecosystem comprises the intestinal microbiome and the intestinal environment, with the intestinal microbiome serving as its central component. The human intestine harbors over 1,000 different types of microorganisms, amounting to a population of more than $1\times10^{14}$ microorganisms [1], which play vital roles in various physiological functions, such as human metabolism, nutrition, immunity, and more. The establishment and disruption of intestinal microecological homeostasis significantly impact human health and diseases. Notably, they exert a considerable influence on the development and progression of numerous systemic diseases, with the digestive system being the most affected.

Children, as a distinct group, possess unique physiological and pathological characteristics. Due to the ongoing development of their intestinal tract, their intestinal microecology is especially sensitive to external influences, such as maternal diet, the mother's psychological state during pregnancy, the mode of delivery, feeding practices, and the utilization of antibiotics. The present literature has provided evidence of a significant correlation between gastrointestinal diseases related to pediatric surgery, such as neonatal necrotizing enterocolitis (NEC), Hirschsprung's disease (HSCR), inflammatory bowel disease (IBD) and intestinal microecology. Gaining a deeper understanding of the interaction mechanism between intestinal microecology and these diseases may further yield valuable insights into the clinical diagnosis and treatment of gastrointestinal diseases, which could facilitate early screening and diagnosis and slow down the progression of related diseases, ultimately improving the quality of life of the affected children.

## 2 Relationship between intestinal microecology and the pathogenesis of gastrointestinal diseases related to pediatric surgery

### 2.1 Involvement of intestinal microecology-related metabolites in the pathogenesis of gastrointestinal diseases related to pediatric surgery

Intestinal microecology-related metabolites act as a crucial link between the intestinal microbiome and disease, influencing the disease process and outcomes by regulating various metabolic pathways. The association between several metabolites and gastrointestinal diseases has become a prominent area of research, focusing on short-chain fatty acids (SCFAs), bile acids, tryptophan, lipopolysaccharide, sphingolipids, and others. Among these, SCFAs and bile acid metabolites stand out as particularly significant. This article primarily focuses on elucidating the vital roles of SCFAs and bile acids in the pathogenesis of gastrointestinal diseases related to pediatric surgery.

#### 2.1.1 Short-chain fatty acids

SCFAs are metabolites formed by the fermentation of undigestible dietary fibers by anaerobic bacteria, such as *Bifidobacterium*, *Bacteroides*, and *Firmicutes*, in the cecum and colon. The three main SCFAs are acetic acid, propionic acid, and butyric acid. These SCFAs play a crucial role in gastrointestinal health and disease, as they contribute to maintaining the mechanical and chemical barriers of the intestine and can influence host immunity and energy status [2]. SCFAs upregulate the expression of various junction proteins in the intestinal epithelium by activating MAPK, inhibiting the MLCK/MLC2 pathway, and phosphorylating PKCβ2, leading to the formation of tight junctions in the epithelium, thereby maintaining its mechanical barrier of the epithelium [3]. SCFAs regulate intestinal functions mainly through the G protein-coupled receptors (GPCRs) activation pathway and the histone deacetylase activity (HDAC) inhibition pathway, which are key in maintaining the intestinal mucosal barrier. By activating GPCRs, SCFAs promote the secretion of intestinal mucus proteins, antimicrobial peptides, and IgA, which helps to preserve the intestinal mucosal chemical barrier and resist pathogen invasion [4]. Furthermore, SCFAs can influence intestinal mucosal immunity by regulating the function of almost all types of immune cells, such as inducing the production of Foxp3 and IL10 in Tregs cells through the HDAC inhibition pathway, leading to the suppression of the intestinal immune response [5], which is crucial for regulating intestinal inflammation. The levels of SCFAs in the intestine are closely associated with the occurrence and development of certain gastrointestinal diseases related to pediatric surgery.

SCFAs play a dose-dependent role in the pathogenesis of NEC, where higher fecal SCFA levels have been shown to associate with increased damage to the intestine and a higher risk of NEC, while low doses of SCFAs, particularly butyric acid, demonstrating protective effects on the intestine [6]. In preterm infants, due to insufficient gastrointestinal motility, immature intestinal barrier function and poor digestion, their intestine is susceptible to bacterial colonization and carbohydrate malabsorption, leading to the accumulation of excessive SCFAs. In addition, immature intestines are more prone to excessive inflammatory responses compared to immune homeostasis [7]. Certain pathogenic bacteria in the intestine, such as *Clostridium perfringens*, *Clostridium difficile*, *Clostridium paraputrificum*, *Clostridium butyricum* and *Klebsiella pneumoniae*, have been shown to produce excess butyric acid through lactose fermentation, which can lead to mucosal damage in NEC by inducing high expression of the inducible nitric oxide synthase gene. Moreover, the overproduction of SCFAs by intestinal pathogenic bacteria may reduce the integrity of the intestinal epithelial cell barrier in NEC and exacerbate inflammatory responses [8]. It has also been suggested that commensal microbiota (e.g., *Bifidobacteria*, *Lactobacillus*) in the intestine can suppress the inflammatory response in the intestine and help regulate the balance of intestinal immune microecology [8].

Children with IBD often have significant alterations in their intestinal microecology. A study from Harvard University revealed that obligate anaerobic bacteria responsible for producing SCFAs were nearly absent in IBD patients [9] and was associated with a shift in the intestinal microbiome towards a pro-inflammatory microbiome, thereby exacerbating host inflammation and worsening their conditions [10]. The decrease in the number of SCFAs can affect Treg cell differentiation and epithelial cell proliferation [11], diminishing the inhibitory effects of pro-inflammatory cytokine expression, decreasing antimicrobial peptide secretion [1,

12], and influencing the expression of intercellular tight junction proteins, leading to intestinal epithelial damage that disrupts intestinal barrier functions. Furthermore, the increase of *Proteobacteria* and the decrease of SCFAs in the intestine of children contribute to the overproduction of IgG against commensal microbiota, which activates macrophages to produce inflammatory cytokines and overstimulates Th1 or Th17 cells, triggering inflammatory reactions [13]. In children with Crohn's disease (CD), the fecal samples show a reduced abundance of various SCFAs-producing bacteria, such as *Anaerostipes*, *Blautia*, *Coprococcus*, *Faecalibacterium*, *Lachnospira*, *Odoribacter*, *Roseburia*, *Ruminococcus*, and *Sutterella* [14]. The sustained decrease in the abundance of these microbiota leads to a significant reduction in the level of SCFAs in these patients. As a result, the ability of these children to regulate inflammation and repair the epithelium is compromised, contributing to the development and progression of the disease.

The reduced content of SCFAs in children with HSCR may be linked to a decrease in the abundance of *Clostridia* and *Bacteroidia*, which are involved in carbohydrate degradation, and a significant decrease in the level of active carbohydrate-active enzymes [15]. Butyrate, one of the SCFAs, is the primary energy source for colonic epithelial cells and exhibits immunomodulatory and anti-inflammatory effects. Reduced butyrate levels are closely linked to the development of inflammation [16]. Mattar et al. [17] found that the decreased SCFAs content in fecal specimens of HSCR patients leads to a significant reduction in mucin MUC-2 content, consequently compromising the mucin barrier on the intestinal surface. The imbalance in intestinal microecology, resulting in reduced SCFAs content, plays a significant role in the occurrence and development of colitis. Hirschsprung-associated enterocolitis (HAEC) is a common complication of HSCR and a leading cause of death in affected children. A multicenter study on HAEC revealed that children with HAEC exhibited a more than fourfold reduction in total fecal SCFAs compared to those without HAEC. The decrease in SCFAs content disrupts intestinal enterocyte homeostasis, contributing to the progression of HSCR towards HAEC [16].

**2.1.2 Bile acids**

Primary bile acids are synthesized in the liver and released into the intestine, where most are reabsorbed through the enterohepatic circulation. A small fraction (1% to 2%) is converted into secondary bile acids by intestinal flora, particularly *Clostridia*. These secondary bile acids possess anti-inflammatory properties and may serve as indirect biomarkers in the mucosal healing process. Animal studies have demonstrated that bile acids have a direct antimicrobial effect and can inhibit the growth of the intestinal microbiome [17]. The interaction between the intestinal microbiome and bile acids is essential to maintain equilibrium. Disruption of this balance can harm the intestinal barrier function and activate inflammatory signaling pathways, leading to the development of intestinal inflammatory diseases [18]. Bile acids act as signaling molecules primarily by activating nuclear receptors, including farnesol X receptor (FXR) and pregnane X receptor (PXR), as well as GPCRs. By activating FXR, bile acids can exert antibacterial effects and help maintain intestinal barrier function by mediating the synthesis of antibacterial substances [19]. FXR can also reduce local intestinal inflammation and immune responses through fibroblast growth factor 15/19. Additionally, secondary bile acids can activate PXR in epithelial cells, leading to the formation of PXR/RXR complexes that interact with NF-κB, which then inhibits TLR4 signaling cascade responses and reduces TLR4 mRNA expression [20], thereby alleviating intestinal inflammation, promoting intestinal cell proliferation and limiting apoptosis. High concentrations of bile acids in the intestine have also been shown to promote the development of intestinal inflammation.

As the lesion in children with NEC worsens, in which TLR4 has been identified as a crucial player, it has been observed that there is a reduction in the relative abundance of *Clostridia* and a decrease in the content of secondary bile acids, with the diminished levels of bile acids weakening the inhibitory effect of PXR on TLR4, leading to an increased inflammatory response in the intestine and promoting the progression of NEC [20].

In children with IBD, there is a reduction in the abundance of *Firmicutes* and *Bacteroidetes* in the intestine, resulting in a decline in secondary bile acids in the intestinal lumen, leading to a diminished protective effect on intestinal epithelial cells. Additionally, the content of 3-OH vulcanized secondary bile acid increases, weakening the anti-inflammatory effect of secondary bile acids and potentially exacerbating IBD [21]. Modified bile acids can activate Treg and Th cells, especially Th17, modulate the immune response by activating receptors like FXRs and GPCRs, reduce the activity of bile salt hydrolases in IBD patients and inhibit or promote inflammation [22]. In the liver, the FXR-FGF15/19 signaling pathway activated by bile acids downregulates the expression of hepatic enzymes involved in bile acid synthesis, resulting in reduced bile acid production. Patients with CD exhibit decreased levels of FGF19, which leads to reduced FXR activation and inhibition of the FXR-FGF15/19 pathway. Consequently, this imbalance promotes an increase in intrahepatic bile acid synthesis, contributing to the development of intestinal inflammation [23].

**2.2 Intestinal microecology plays an immune role in the pathogenesis of gastrointestinal diseases related to pediatric surgery**

Intestinal microecology plays a vital role in the intestinal immune system as it is actively involved in maintaining both natural and adaptive immunity. The innate immune response functions through pattern recognition receptors, particularly Toll-like receptors (TLRs) and nod-like receptors (NLRs). TLR4 and NLRs are expressed in intestinal epithelial cells, with TLR4 regulating apoptosis, migration and proliferation of these cells. TLRs recognize conserved structural molecules of invading microorganisms, triggering immune cell responses and inflammation, while NLRs can identify invading pathogenic microorganisms. In the context of adaptive immune responses, the intestinal microbiome significantly influences the development of B cells. Then, the microbial complex-sIgA exerts anti-inflammatory effects by down-regulating the production of pro-inflammatory cytokines, such as IL-8. Th17 cells are an essential component of the intestinal mucosal immune system that produce pro-inflammatory factors [24].

TLR4 plays a crucial role in the pathogenesis of NEC by inducing intestinal injury and inhibiting intestinal mucosal repair in premature infants. Children with NEC have significantly higher levels of TLR4 mRNA and protein in their intestinal epithelial cells than those without the disease. An animal experiment demonstrated that mice lacking TLR4 in their intestinal epithelial cells did not develop NEC [25]. In children with NEC, the presence of *Proteobacterias* has been shown to trigger a cascade of inflammatory and immune responses by activating the TLR4 signaling pathway, resulting in the disruption of the integrity of the intestinal mucosal barrier and the release of pro-inflammatory factors. There are abundant inflammatory Th17 cells in the intestines of NEC patients, and their infiltration into the intestinal tract of newborns is dependent on the TLR4 signaling pathway and has been associated with intestinal inflammation [25]. Moreover, the TLR4 signal in the endothelium of the intestinal mesentery can reduce the expression of human endothelial nitric oxide synthase, leading to decreased nitric oxide content and resulting in mesenteric vasoconstriction, poor blood circulation, and increased apoptosis of intestinal epithelial cells due to local ischemia [26].

The intricate interplay between intestinal mucosal immune mechanisms and the intestinal microbiome is vital for the onset and perpetuation of inflammation in IBD. Changes in the intestinal microecology in individuals with genetic susceptibility can trigger an abnormal mucosal immune response [11], resulting in chronic inflammation in the intestine.The rise in pathogenic bacteria in the intestinal tract leads to a disrupted microecological balance, which damages the intestinal mechanical and immune barriers through invasiveness and toxin secretion, consequently causing an imbalance in the immune response, inflammation, and increased apoptosis. The disrupted intestinal microecology also prompts lymphocytes in mesenteric lymph nodes to secrete numerous inflammatory factors, triggering inflammatory response, which results in tissue damage. In patients with CD, the main manifestations include increased levels of IFN-γ, TNF-α and IL17, while ulcerative colitis (UC) is characterized by elevated TNF-α, IL-5 and IFN-γ levels [27]. Disruptions in intestinal microecology during the neonatal period can result in increased pro-inflammatory cytokines (IL-12, TNF-γ), which can activate Th1, damage the immune system and elevate the risk of IBD [11]. It has been observed that patients with CD exhibit a significant increase in the number of Th17 cells and their associated cytokines [28], which has been positively correlated with the occurrence and progression of IBD. Th17 cells promote the recruitment of inflammatory cells by activating CXC chemokines and inhibit the negative regulation of Treg cells, thus eliciting a robust inflammatory response [11].

## 3 The application of intestinal microecology in the diagnosis of gastrointestinal diseases related to pediatric surgery

### 3.1 The role of intestinal microecological diversity in the diagnosis of gastrointestinal diseases related to pediatric surgery

In certain gastrointestinal diseases related to pediatric surgery, the composition of the intestinal microbiome undergoes a series of alterations compared to that of healthy children, and these changes vary at different stages of the disease. Advances in sequencing technology, bioinformatic analysis and the decreasing cost of sequencing have facilitated the use of fecal sample microbiome testing as a non-invasive diagnostic tool, offering novel approaches for the diagnosis of gastrointestinal diseases related to pediatric surgery.

Early changes in the intestinal microbiome of children with NEC can be used to predict the risk of disease development. Olm et al. [29] demonstrated that the rate of bacterial growth was the most robust predictor of NEC by analyzing changes in the intestinal microbiome of 160 children with NEC and healthy children. Furthermore, in a meta-analysis conducted by Tarracchini et al. [30], the researchers reported early elevated concentrations of *Clostridium neonatale* and *Clostridium perfringens* species in children with NEC. In early-onset NEC cases, there was mainly an increased abundance of *Clostridium sensu stricto* before disease onset [31], while late-onset NEC showed increased *Gammaproteobacteria* and *Escherichia/Shigella* [32]. Moreover, an increased predominance of *Citrobacter koseri, Clostridium perfringens* [30] and/or *Klebsiella pneumoniae*, along with a decrease in intestinal microecological diversity and *Lactobacillus* abundance, have been associated with NEC risks in premature infants [33], while stable intestinal microbiomes and increased levels of *Bifidobacteria* was found to associate with reduced NEC risks [34]. These changes in the microbiome hold potential as microbial markers for the early diagnosis of NEC and could contribute to its early prevention and diagnosis.

The diversity, abundance and uniformity of intestinal microecology in children with IBD are lower than in normal children, and these changes in intestinal microecology hold significant diagnostic implications for the disease. Specific alterations have been observed in the intestinal microbiome of children with IBD, including increased levels of *Escherichia/Shigella* and *Bifidobacteriaceae* and decreased levels of *Bifidobacteria* and *Veillonellaceae*. By combining these changes, predictive models have shown high sensitivity and specificity for diagnosing IBD in children [35]. Furthermore, a decrease in *Roseburia* abundance in feces may precede the onset of IBD [36]. The activity of IBD was shown to be positively correlated with *Proteobacterias* and *Bacillus* and negatively correlated with *Clostridia* and *Actinomycetes* [37]. The severity of IBD was negatively correlated with the main bacteria producing SCFAs, such as *Firmicutes*, *Roseburia*, *Ruminococcus* and *Blautia*, but positively correlated with some sulfide-producing bacteria, such as *Fusobacterium*, *Veillonella*, *Atopobium*, *Streptococcus* and *Leptotrichia* [38]. These provide new insights for the early diagnosis and assessment of disease severity. A multicenter study found that *Pasteurellaceae, Veillonellaceae*, *Neisseriaceae* and *Fusobacteriaceae* were positively correlated with the abundance of CD in children. *Bacteroides*, *Faecalibacterium*, *Roseburia*, *Blautia*, *Ruminococcus, Coprococcus* and several other taxa within the families of *Ruminococcaceae* and *Lachnospiraceae* are negatively correlated with CD and could be used as important biomarkers for disease assessment [39]. Additionally, the lack of *Bifidobacterium* in feces can be used as a biological marker for the characteristic intestinal microecological imbalance observed in CD [36].

A prospective multicenter study on HAEC reported that *Flavonifractor plautii* and *Eggerthella lenta*, and to a lesser extent *Ruminococcus gnavus*, in feces were closely related to the occurrence of HAEC in children aged 0-2 years old [40] and could be considered as a marker for early diagnosis of HAEC in HSCR.

**3.2 Diagnostic role of fecal volatile metabolites in gastrointestinal diseases related to pediatric surgery**

Fecal volatile organic compounds (VOCs) have shown promise as a non-invasive marker for early disease diagnosis, reflecting the composition of the intestinal microbiome.

Studies have revealed that certain VOCs in the intestine can serve as potential markers for diagnosing NEC. For instance, 2-3 days before NEC onset, VOCs could distinguish between NEC and non-NEC children with a sensitivity of 83% and specificity of 75% [41]. In a prospective, multicenter study in the UK, Probert et al. observed changes in VOCs, by-products of bacterial metabolism, 3-4 days before NEC onset [42]. An increase in specific VOCs, such as methyl-aldehydes, 3-methylbutanal and 2-methylbutanal, was associated with an increased risk of NEC. On the other hand, VOCs such as butane-2-one, carbon disulfide, 3-methyl and 2-methylbutanoic were found to reflect the abundance of *Bifidobacteria* in the gut, and their increase was associated with a reduced risk of NEC [42].

In a prospective, dual-center case-control study [43] and a prospective, multicenter case-control study [44] in the Netherlands, it was found that fecal VOCs in children with IBD had higher diagnostic value as a non-invasive biomarker, and its specificity was significantly higher than that of fecal calcitonin. Fecal calcitonin relies on its high sensitivity to mucosal inflammation (97%) for diagnosing IBD, but its specificity in children with IBD is only 70% [45], while the sensitivity and specificity of VOCs in IBD diagnosis was 86% and 85% [46]. Although detecting fecal VOCs in children with suspected IBD may help reduce the number of endoscopic examinations required, further research is needed to explore the differential diagnosis of UC and CD using fecal VOCs.

## 4 Application of intestinal microecology in the treatment of gastrointestinal diseases related to pediatric surgery

### 4.1 Role of microecological agents in the treatment of gastrointestinal diseases related to pediatric surgery

Microecological agents include probiotics, prebiotics, and synbiotics. These agents can regulate intestinal microecological balance by influencing the intestinal microbiome and metabolites, enhancing intestinal barrier function, regulating intestinal mucosal immune function and modulating nutrient metabolism. Additionally, genetically engineered bacteria offer a more precise approach to disease treatment using microecological agents. Previous studies have demonstrated the therapeutic efficacy of microecological agents in managing gastrointestinal diseases related to pediatric surgery.

Probiotics/prebiotics can increase the abundance of SCFAs and SCFAs-producing bacteria in premature infants, possess anti-inflammatory effects in

NEC and protect fetal intestinal epithelial cells from pro-inflammatory cytokines and chemokines by inhibiting different cellular signaling pathways [8]. Samantha et al. [47] suggested that the current results of randomized controlled trials on the use of single-strain probiotics in premature infants are not convincing for probiotics to reduce the incidence of NEC, while the combined use of multiple strains of probiotics may play a certain role in reducing the incidence of NEC. A multicenter trial in Australia and New Zealand showed that the combination of *B infantis*, *Streptococcus thermophilus* and *B lactis* reduced the incidence of NEC in infants (<1500 g) (4.4 down to 2%) [48]. Despite the potential benefits, the safety of oral probiotics in premature infants needs to be further investigated. Premature infants have an immature immune system and higher intestinal mucosal permeability, making them susceptible to translocation of probiotics from the gut to the bloodstream, which could lead to systemic infection.

The treatment of IBD involves inducing remission and maintaining remission. Probiotics have been shown to be beneficial in maintaining remission and reducing the recurrence rate of IBD, but the optimal choice, dosage, administration mode and duration of probiotics require further investigation. VSL#3, a mixture of 8 beneficial bacteria, has been recommended by the Yale/Harvard consensus and the European Society for Clinical Nutrition and Metabolism for both IBD maintenance and induced remission [49]. While 5-aminosalicylic acid is the first-line treatment for mild to moderate UC-induced remission, probiotics can be considered as an alternative for mild UC cases where 5-aminosalicylic acid is not tolerated. The use of VSL#3 in UC children has shown a higher rate of induced clinical remission than routine treatment, with a lower recurrence rate within one year [50]. However, caution should be exercised in using probiotics in children with severe acute UC, as the damaged intestinal barrier may allow probiotics to enter the bloodstream, resulting in bacteremia [51]. The use of probiotics in patients with CD remains controversial. Fedorak et al. [52] showed no significant difference in the prevention of recurrence between VSL#3 and the placebo group for CD. On the other hand, Gupta et al. [53] showed that Lactobacillus rhamnosus GG (LGG) significantly improved the clinical symptoms of children with mild to moderate active CD. However, Bousvaros et al. [54] showed that adding LGG to standard treatment did not affect remission in children with CD. Apart from probiotics, probiotics may play a certain role in the treatment of children with IBD. The protective effects of probiotics in IBD treatment are mostly focused on inulin and oligosaccharides [55]. Inulin can increase the number of *Lactobacillus*, reduce intestinal PH, increase butyric acid synthesis and inhibit inflammatory damage to intestinal epithelial cells. Galacto-oligosaccharide can enrich *Bifidobacteria* and *Lactobacillus* in the large intestine, prevent pathogenic bacteria from infection, and increase the content of SCFAs [56]. Currently, there is limited experimental data on the treatment of IBD with probiotics and synbiotics in children, and their use in the induced remission of IBD in children is not recommended. However, the latest research suggests that engineered bacteria expressing anti-inflammatory metabolites or proteins, inhibiting inflammation, and regulating intestinal flora may have a role in preventing and treating IBD [57].

Microecological agents, such as Live Combined *Bifidobacterium* and *Lactobacillus* tablets, may help prevent HAEC after HSCR surgery. They have been shown to effectively reduce the levels of inflammatory factors in children, thereby reducing mucosal damage, inhibiting the proliferation of pathogenic bacteria, and effectively reducing the incidence of HAEC in these children [58].

Currently, the studies on intestinal microecological agents for the treatment of short bowel syndrome (SBS) primarily involve animal experiments, and further clinical studies are needed to validate their efficacy. In an experiment involving SBS rats, it was found that they exhibited increased amplitude and frequency of jejunum and ileum contractions to heightened intestinal peristalsis, leading to diarrhea, and nutritional malabsorption. However, probiotics (specifically *Lactobacillus* and *Bifidobacteria*) effectively regulate intestinal peristalsis, restoring it to normal levels and improving nutrient absorption in SBS [59]. In a piglet experiment, the combined administration of *Lactobacillus* and *Bifidobacteria* resulted in an increased relative abundance of probiotics in the mucous membrane and lumen microflora, greater microbial diversity, elevated concentrations of SCFAs in fecal and colonic effluents, and minimal risk of septicemia associated with *Lactobacillus* and *Bifidobacteria* [60]. In a study involving 18 children with SBS, no significant differences in fecal microbiome or overall growth were observed between those who received *Lactobacillus rhamnosus* and *Lactobacillus johnsonii* versus the placebo group [61].

**4.2 Role of fecal microbiota transplantation in the treatment of gastrointestinal diseases related to pediatric surgery**

FMT involves transferring functional microflora from healthy individuals' feces into the intestinal tract of patients to reconstruct a new intestinal microbiome. FMT has therapeutic potential by restoring intestinal microecological balance, reducing inflammation, enhancing immune system function, and restoring intestinal barrier integrity. Due to the ongoing development of children's gut microbiota and immune systems, they may be more sensitive to FMT.

Currently, research on FMT in NEC is limited to animal models, lacking clinical experimental results. However, in animal studies, FMT was shown to successfully reverse the severity and slow the progression of NEC in mice. It also effectively promoted the re-establishment of the intestinal microecological structure, leading to a quick increase in intestinal microbial diversity and abundance to previous levels. Furthermore, FMT in mice can eliminate oxidative stress, inhibit TLR4-mediated pro-inflammatory signals in the intestines, reduce intestinal injury and inflammation in the terminal ileum, and restore the intestinal mucosal barrier and immune molecular network [62]. In terms of drug administration, findings from animal experimental studies indicate that oral FMT alone may exacerbate intestinal injury [63, 64]. On the other hand, rectal FMT alone can reduce the incidence of NEC, while combined oral and rectal FMT can increase the total mortality of NEC. Transrectal FMT has been shown to down-regulate inflammatory responses and promote intestinal tight junctions [64]. Moreover, rectal administration reduces the exposure time of potentially pathogenic microorganisms in the proximal intestine, resulting in a decreased incidence of fatal septicemia and overall mortality related to FMT [65], which is superior to oral and combined FMT.

In the treatment of IBD in children, FMT has the potential to play a therapeutic role by increasing the α diversity, regulating intestinal microflora, promoting the growth of SCFAs-producing bacteria, increasing the content of SCFAs, reducing the levels of inflammatory cytokines such as TNF-α and IL-6, inhibiting of inflammatory responses and modulate of host immunity [66]. Gerasimidis et al. [34] showed that in 6 studies comprising children with UC and two studies involving pediatric CD, the cumulative remission rates in children with UC (n=34) and CD (n=13) were 54% and 23%, respectively. In a single-center prospective study, Alka-Goyal et al. reported that after six months of follow-up, 28% (6/21 cases) of the patients achieved clinical remission with a single FMT, indicating short-term effectiveness and relative safety in young patients with active IBD [67]. The first international expert consensus on FMT in the treatment of IBD indicates that serious adverse reactions in FMT for IBD are rare, but if occurred, they could be mainly related to the mode of administration [68]. FMT administered via nasal duodenal tube and enema has shown effectiveness in achieving clinical remission of IBD, with a lower incidence of adverse events for the latter approach [68]. However, the optimal timing, dosage, and treatment duration of FMT for IBD are still unclear and require further investigation.

The use of FMT in the treatment of SBS in children has been relatively less reported. In 2017 and 2018, two cases of SBS secondary to D-lactic acidosis were successfully treated with FMT, as reported by Davidovics et al. [69] and Bulik-Sullivan et al. [70]. In an animal study involving piglets with SBS, FMT was found to increase the diversity of intestinal microecology in the short term without any significant increase in the risk of sepsis associated with FMT [71].

Children and adolescents with lymphoma may benefit from immune checkpoint inhibitors (ICIs). While ICIs have been shown to effectively improve the prognosis of patients with malignant tumors, they can also lead to immune-related adverse reactions, with immune-mediated colitis (IMC) being one of the most common and serious reactions. Multiple reports [72] suggest that the intestinal flora plays a role in regulating ICIs, and patients with diverse intestinal flora may experience better immunotherapy effects. FMT contributes to the reconstruction of the microbiota. In this regard, animal experiments and adult clinical trials [73] have demonstrated that FMT may inhibit the occurrence and development of ICIs-related colitis, increase the diversity of α-intestinal flora in IMC patients, alter mucosal immune infiltration, reduce the density of CD8+T cells, and increase the density of Treg cells in the colonic mucosa. It can also serve as an alternative treatment for IMC. However, there is currently a lack of clinical research on the application of FMT in children with tumor ICSs-related IMC.

The application of FMT in disease treatment is influenced by various factors, including the source of the donor, the condition of the recipient, the FMT regimen (i.e., fecal volume and infusion frequency, route of administration, and adjuvant therapy), and other related factors. Currently, there is relatively limited clinical research on FMT in gastrointestinal diseases related to pediatric surgery, and several challenges persist in its application for disease treatment, such as the unclear composition of

FMT treatment, the exact mechanism of action being unknown, the selection of appropriate recipients and donors, and uncertainty regarding its long-term safety and effectiveness [74]. Thus, large-scale multicenter prospective randomized controlled trials are still required to thoroughly evaluate the efficacy and safety of FMT in treating pediatric surgery-related gastrointestinal diseases.

**Conclusion and outlook**

Intestinal microecology is an important aspect of the development and progression of pediatric surgery-related gastrointestinal diseases. However, its interaction with specific diseases remains unclear, and further evidence from animal models and clinical studies is needed, with translational research findings from animal experiments to clinical applications being a key area for future investigation. The use of fecal microflora in disease diagnosis still has some limitations. Fecal samples may not fully represent the diverse contents of different regions of the intestinal tract simultaneously, and our understanding of the distribution of microorganisms, their internal environment, and biochemical activities in the intestine is still limited. In this regard, research and development on sampling capsule equipment could provide potential solutions. Furthermore, the application of intestinal microecology in disease diagnosis and treatment lacks standardized understanding and guidelines, urging the need for more extensive and in-depth research, along with a comprehensive collection of clinical data and evidence to establish a consensus among experts and regulate its clinical use. As research on intestinal microecology continues to progress, it is expected that innovations in this field will increasingly contribute to the diagnosis, treatment and prevention of pediatric surgery-related gastrointestinal diseases, ultimately providing more effective and targeted approaches that may benefit children's health and well-being.

**Author contributions**

All authors have read and agreed to the published version of the manuscript.

**Statement of financial support**

This research did not receive any specific grant from funding agencies in the public, commercial, or not-for-profit sectors.

**Declaration of competing interest**

The authors declare that they have no known competing financial interests or personal relationships that could have appeared to influence the work reported in this paper.